%% file: 3I.tex
\documentclass[]{aastex7}
\usepackage{ulem}
\usepackage{mhchem}
\usepackage{bm}

\newcommand\subs[1]{\textsubscript{#1}}

\newcommand\rh[1]{\textcolor{black}{{\textit{r}\subs{\textit{H}}}#1}}

\newcommand\trot[1]{\textcolor{black}{{\textit{T}\subs{rot}}#1}}

\newcommand\kms[1]{\textcolor{black}{{km\,s$^{-1}$}#1}}
\newcommand\ps[1]{\textcolor{black}{{s$^{-1}$}#1}}

\definecolor{gold}{rgb}{0.64,0.54,0.29}

\received{2026 March 20}
\revised{2026 May 8}
\accepted{}
\submitjournal{The Astrophysical Journal Letters}

\shorttitle{3I/ATLAS Coma Physics}
\shortauthors{Roth et al.}

\begin{document}


\title{Coma Physics of an Interstellar Object: JWST Spatial-Spectral Mapping of 3I/ATLAS}

\correspondingauthor{Nathan X. Roth}
\email{nathaniel.x.roth@nasa.gov}

\author[0000-0002-6006-9574]{Nathan X. Roth}
\affiliation{Solar System Exploration Division, NASA Goddard Space Flight Center, 8800 Greenbelt Rd, Greenbelt, MD 20771, USA}
\affiliation{Department of Physics, American University, 4400 Massachusetts Ave NW, Washington, DC 20016, USA}
\email{nathaniel.x.roth@nasa.gov}

\author[0000-0001-8233-2436]{Martin A. Cordiner}
\affiliation{Solar System Exploration Division, NASA Goddard Space Flight Center, 8800 Greenbelt Rd, Greenbelt, MD 20771, USA}
\affiliation{Department of Physics, The Catholic University of America, 620 Michigan Ave., N.E. Washington, DC 20064, USA}
\email{martin.cordiner@nasa.gov}

\author[0000-0001-7694-4129]{Stefanie N. Milam}
\affiliation{Solar System Exploration Division, NASA Goddard Space Flight Center, 8800 Greenbelt Rd, Greenbelt, MD 20771, USA}
\email{stefanie.n.milam@nasa.gov}

\author[0000-0001-6752-5109]{Geronimo L. Villanueva}
\affiliation{Solar System Exploration Division, NASA Goddard Space Flight Center, 8800 Greenbelt Rd, Greenbelt, MD 20771, USA}
\email{geronimo.l.villanueva@nasa.gov}

\author[0000-0001-6752-5109]{Steven B. Charnley}
\affiliation{Solar System Exploration Division, NASA Goddard Space Flight Center, 8800 Greenbelt Rd, Greenbelt, MD 20771, USA}
\email{steven.b.charnley@nasa.gov}

\author[0000-0003-2414-5370]{Nicolas Biver}
\affiliation{LIRA, Observatoire de Paris, Université PSL, CNRS, Sorbonne Université, Université Paris Cité, 5 place Jules Janssen, 92195 Meudon, France}
\email{nicolas.biver@obspm.fr}

\author{Dominique Bockelée-Morvan}
\affiliation{LIRA, Observatoire de Paris, Université PSL, CNRS, Sorbonne Université, Université Paris Cité, 5 place Jules Janssen, 92195 Meudon, France}
\email{dominique.bockelee@obspm.fr}

\author[0000-0002-2668-7248]{Dennis Bodewits}
\affiliation{Physics Department, Edmund C. Leach Science Center, Auburn University, Auburn, AL 36849, USA}
\email{dennis@auburn.edu}

\author[0000-0003-2110-8152]{Steven J. Bromley}
\affiliation{Physics Department, Edmund C. Leach Science Center, Auburn University, Auburn, AL 36849, USA}
\email{sjb0068@auburn.edu}

\author{Jacques Crovisier}
\affiliation{LIRA, Observatoire de Paris, Université PSL, CNRS, Sorbonne Université, Université Paris Cité, 5 place Jules Janssen, 92195 Meudon, France}
\email{jacques.crovisier@obspm.fr}

\author[0000-0001-7479-4948]{Maria N. Drozdovskaya}
\affiliation{Physikalisch-Meteorologisches Observatorium Davos und Weltstrahlungszentrum (PMOD/WRC), Dorfstrasse 33, CH-7260, Davos Dorf, Switzerland}
\email{maria.drozdovskaya.space@gmail.com}

\author[0000-0003-0194-5615]{Sara Faggi}
\affiliation{Solar System Exploration Division, NASA Goddard Space Flight Center, 8800 Greenbelt Rd, Greenbelt, MD 20771, USA}
\affiliation{Department of Physics, American University, 4400 Massachusetts Ave NW, Washington, DC 20016, USA}
\email{sara.faggi@nasa.gov}

\author[0000-0003-0774-884X]{Davide Farnocchia}
\affiliation{Jet Propulsion Laboratory, California Institute of Technology, 4800 Oak Grove Dr., Pasadena, CA 91109, USA}
\email{Davide.Farnocchia@jpl.nasa.gov}

\author[0000-0002-2026-8157]{Kenji Furuya}
\affiliation{Pioneering Research Institute, RIKEN, 2-1 Hirosawa, Wako-shi, Saitama, 351-0198, Japan}
\email{kenji.furuya@riken.jp}

\author[0000-0002-6702-7676]{Michael S. P. Kelley}
\affiliation{Department of Astronomy, University of Maryland, College Park, MD 20742-0001, USA}
\email{msk@astro.umd.edu}

\author[0000-0001-7895-8209]{Marco Micheli}
\affiliation{ESA NEO Coordination Centre, Planetary Defence Office, European Space Agency, Largo Galileo Galilei, 1, 00044 Frascati (RM), Italy}
\email{marco.bs.it@gmail.com}

\author[0000-0003-2152-6987]{John W. Noonan}
\affiliation{Physics Department, Edmund C. Leach Science Center, Auburn University, Auburn, AL 36849, USA}
\email{noonan@auburn.edu}

\author[0000-0001-7694-4129]{Cyrielle Opitom}
\affiliation{Institute for Astronomy, University of Edinburgh, Royal Observatory, Edinburgh EH9 3HJ, UK}
\email{copi@roe.ac.uk}

\author[0000-0003-4365-1455]{Megan E. Schwamb}
\affiliation{Astrophysics Research Centre, School of Mathematics and Physics, Queen's University Belfast, Belfast BT7 1NN, UK}
\email{m.schwamb@qub.ac.uk}

\author[0000-0003-3091-5757]{Cristina A. Thomas}
\affiliation{Northern Arizona University, Department of Astronomy and Planetary Science,
P.O. Box 6010, Flagstaff, AZ, 86011 USA}
\email{cristina.thomas@nau.edu}




\input{Abstract}



\input{Intro}

\input{Observations}
\input{Results}
\input{Discussion}
\input{Conclusion}

\begin{acknowledgments}
This work is based on observations made with the NASA/ESA/CSA James Webb Space Telescope. The data were obtained from the Mikulski Archive for Space Telescopes at the Space Telescope Science Institute, which is operated by the Association of Universities for Research in Astronomy, Inc., under NASA contract NAS 5-03127 for JWST. The specific observations analyzed can be accessed via \dataset[doi: 10.17909/0ek5-h758]{https://doi.org/10.17909/0ek5-h758}. Analysis was supported via STScI grant JWST-GO-05094.001. We gratefully acknowledge the assistance of optical observers who submitted astrometric observations of 3I/ATLAS in the weeks leading up to our observations, to help refine the ephemeris position. In particular, we thank J. Chatelain, E. Gomez, S. Greenstreet, W. Hoogendam, C. Holt, H. W. Lin, T. Lister, T. Santana-Ros, L. Salazar Manzano, D. Seligman, Q. Ye, Q. Zhang, K. Meech, and C. Chandler. Supporting astrometric observations were obtained by the Comet Chasers school outreach program (https://www.cometchasers.org/), led by Helen Usher, which is funded by the UK Science and Technology Facilities Council (via the DeepSpace2DeepImpact Project), the Open University and Cardiff University. It accesses the LCOGT telescopes through the Schools Observatory/Faulkes Telescope Project (TSO2025A-00 DFET-The Schools’ Observatory), which is partly funded by the Dill Faulkes Educational Trust, and through the LCO Global Sky Partners Programme (LCOEPO2023B-013). Observers included individuals and representatives from the following schools and clubs: E. Maciulis, A. Bankole, J. Bower, O. Roberts, participants on the British Astronomical Associations’ Work Experience project 2025 from The Coopers Company \& Coborn School; Upminster, UK;  St Marys Catholic Primary School, Bridgend, UK; J. M. Perez Redondo \& Students: A. Matea, L. Guillamet, A. Montoy, and A. Martin from Institut d’Alcarràs, Catalonia, Spain; Louis Cruis Astronomy Club, Brazil; Jelkovec High School, Zagreb, Croatia, and C. Wells at a British Astronomical Association event. This research has made use of NASA’s Astrophysics Data System Bibliographic Services. This research has made use of data and/or services provided by the International Astronomical Union's Minor Planet Center. N.X.R., M.A.C., S.B.C., and S.N.M. were supported by the NASA Planetary Science Division Internal Scientist Funding Program through the Fundamental Laboratory Research work package (FLaRe). M.E.S. acknowledges support in part from UK Science and Technology Facilities Council (STFC) grant ST/X001253/1.
D.F. conducted this research at the Jet Propulsion Laboratory, California Institute of Technology, under a contract with the National Aeronautics and Space Administration (80NM0018D0004). We thank an anonmyous referee for their feedback, which we feel improved the manuscript.
\end{acknowledgments}

\software{Astropy \citep{astropy:2013, astropy:2018, astropy:2022},
Astroquery \citep{Ginsburg2019},
photutils \citep{bradley2025},
jwstComet \citep{Roth2026b}
}


\appendix
\input{obsAppendix}
\input{waterAppendix}
\input{contAppendix}
\input{opacityAppendix}
\input{ModelingAppendix}

\input{sublimeAppendix}
\input{hcnAppendix}

\clearpage

\bibliography{3I}{}
\bibliographystyle{aasjournalv7}



\end{document}

%% file: Abstract.tex
\begin{abstract}

We report a survey of molecular emission from cometary volatiles using the James Webb Space Telescope (JWST) toward interstellar object 3I/ATLAS carried out on UT 2025 December 22 and 23 at a heliocentric distance (\rh{}) of $2.37-2.41$ au. These measurements of CO, \ce{CO2}, \ce{H2O}, \ce{CH3OH}, and \ce{CH4} sampled molecular chemistry in 3I/ATLAS as it receded from its encounter with our Sun and entered the vicinity of the \ce{H2O} ice line --- the region between \rh{} = $2-3$ au where the temperature becomes too low for H$_2$O to vigorously sublime and CO and \ce{CO2} begin to control the overall activity. CO was the most abundant molecule, followed by \ce{H2O} and \ce{CO2}, whose molecular abundances with respect to CO were $(40.5\pm3.1)\%$ and ($41.6\pm0.3)\%$, respectively. This work presents spatial-spectral maps of column density and rotational temperature as a function of distance from the nucleus for all detected species. The spatial distributions of both quantities were highly anisotropic for the apolar species in the coma of 3I/ATLAS, yet were more nearly symmetric for the polar molecules. These results demonstrate how volatiles were segregated in the nucleus ices of 3I/ATLAS and reveal heating and cooling mechanisms in its coma. Derived maps of the ortho-to-para ratio (OPR) for \ce{H2O} were flat with increasing distance from the nucleus and consistent with a coma-averaged value $\mathrm{OPR}=2.7\pm0.2$, slightly less than the expected equilibrium value of three. 

\end{abstract}

%% file: Intro.tex
\section{Introduction} \label{sec:intro}
Comets provide a window into the early solar system. They formed in the cold disk midplane of the protosolar disk during the era of planet formation and were subsequently scattered to the Kuiper disk or the Oort cloud. Studying the volatile composition of their nuclei gives clues to the chemistry and prevailing conditions present where and when they formed in the disk \citep{Mumma2011a,Biver2024b}. Alongside studies of overall composition, near-infrared spectroscopy of comets can provide insights into the physics of their comae, or expanding atmospheres of gas and dust, by sampling ro-vibrational transitions of a suite of volatiles. Such works, both observational and theoretical, have traditionally focused on solar system comets which were measured at \rh{}$<2$ au, where \ce{H2O} is the most abundant coma molecule and drives outgassing and overall thermal physics \citep[e.g.,][]{Tenishev2008,Fougere2012,Bonev2013,Bonev2014}. 

Here we report a study of the coma physics and chemistry of an altogether different subject, the interstellar object 3I/ATLAS, measured at \rh{}$\sim2.4$ au post-perihelion using the JWST NIRSpec integral field unit (IFU) covering $\lambda=1.06-5.14$ $\mu$m. The third interstellar object discovered to date \citep[and the second with a comet-like coma;][]{Seligman2025}, the coma composition of 3I/ATLAS throughout its perihelion passage has been reported with spectroscopy conducted across the electromagnetic spectrum \citep[e.g.,][]{Biver2026,Coulson2026,Hutsemekers2026,Roth2026a,Cordiner2025b,Xing2025,Hoogendam2025,Alvarez2025}. Our post-perihelion JWST observations afforded the opportunity to leverage the spatial-spectral mapping capabilities of the NIRSpec IFU to conduct high-resolution spaxel-by-spaxel studies of the column density and rotational temperature distributions of multiple species in its coma (each spaxel subtends $0\farcs1$, corresponding to $\sim130$ km projected distance per spaxel at the distance of 3I/ATLAS from JWST). These species were CO, \ce{CO2}, and \ce{H2O}, which are often the most abundant molecules in cometary nuclei and take turns driving activity depending on \rh{} \citep[e.g.,][]{Harrington2022}, as well as the trace species \ce{CH3OH} and \ce{CH4}. Section~\ref{sec:obs} provides details of the observations and data analysis procedures. Section~\ref{sec:results} presents our results, and Section~\ref{sec:discussion} discusses our results in the context of solar system comets studied to date.

%% file: Observations.tex
\section{Observations and Data Reduction} \label{sec:obs}
Interstellar object 3I/ATLAS reached perihelion ($q=1.35$ au) on 2025 October 29. We conducted post-perihelion observations of 3I/ATLAS using the JWST NIRSpec IFU. The object's \rh{} ranged from $2.37-2.41$ au, its distance from the telescope ($\Delta_{\mathrm{JWST}}$) from $1.79-1.80$ au, and the solar phase angle ($\phi$) ranged from $22.7\degr-21.6\degr$. Observations used a representative resolving power $\lambda/\Delta\lambda\sim2700$. A single 642 s exposure of the G235H/F170LP grating starting on UT 2025 December 22 03:36 sampled emission from $\lambda=1.06-3.05$ $\mu$m and captured the \ce{H2O} $\nu_1+\nu_3$ vibrational bands near 2.68 $\mu$m. Five 700 s exposures of the G395H/F290LP grating beginning on UT 2025 December 23 08:07 characterized emission from $\lambda=2.87-5.14$ $\mu$m. This setting measured molecular emission from CO $(v=1-0)$, \ce{CO2} $(\nu_3)$, \ce{CH3OH} $(\nu_2,\nu_3,\nu_9)$, \ce{CH4} $(\nu_3)$, and \ce{H2O} hot bands near 4.5 $\mu$m. 

Background exposures with identical circumstances (exposure time, grating settings) were planned offset from the comet position by $300''$; however, two of the five background exposures for the G395H grating failed owing to background star issues, with follow-up attempts scheduled for later in 2026. Careful examination of the available background exposures did not reveal emission from any known interstellar infrared sources or zodiacal light, so analysis proceeded without background subtraction to maximize the signal-to-noise ratio (SNR). Exposures were processed using the JWST Pipeline version 1.20.2 with CRDS jwst\_1464.pmap context files and aligned onto a common spatial-spectral axis using the Drizzle algorithm \citep{Law2023}. An observing log is provided in Appendix~\ref{sec:obslog}.

We generated spatial-spectral maps of molecular column density ($N$, m$^{-2}$) and rotational temperature (\trot{}, K) by performing a spaxel-by-spaxel analysis of the NIRSpec IFU data cubes. Each modeled quantity was retrieved using molecular emission models in the NASA Planetary Spectrum Generator \citep[PSG;][]{Villanueva2018}. We extracted and modeled spectra from each $0\farcs1$ spaxel using the \texttt{jwstComet} package \citep{Roth2026b}, which provides for flexible spectral extraction from JWST IFU data cubes using functions from the \texttt{astropy} and \texttt{photutils} libraries, followed by automated interfacing with the NASA PSG API. The data cubes were converted from units of MJy sr$^{-1}$ to Jy pixel$^{-1}$, then spectra were extracted from the data cubes using the \texttt{photutils} RectangularAperture function in combination with the aperture\_photometry function (method = `center'). Fluxes and $1\sigma$ instrumental noise on a per-spaxel basis were derived from the \texttt{SCI} and \texttt{ERR} extensions of the FITS files.

Contributions from gaseous and continuum emission were identified by comparing the spectra with expected spectral line positions for each species from quantum mechanical models of fluorescent emission generated with the PSG. Each spectrum was baseline subtracted using second- or third-order polynomial baselines. This baseline accounts for continuum emission from the dust and nucleus, as well as scattered sunlight and instrumental artifacts. We chose the lowest possible polynomial order that could account for the spectral shape while avoiding higher order polynomials to avoid introducing spurious features into the spectra. The baselines were fit simultaneously with the molecular emission models using the Optimal Estimation Method implemented in the PSG. Techniques employing simultaneous fitting of the continuum baseline and molecular emission models have been applied to decades of cometary infrared spectroscopy studies \citep[e.g.,][]{DiSanti2003,Villanueva2011a,Bonev2014,DiSanti2017,Roth2018,Faggi2019,Ejeta2024}. Fitting the baselines and emission models simultaneously ensured that uncertainties on the baseline fit were propagated into uncertainties on each retrieved quantity (i.e., $N$ or \trot{}). These fits included a correction for opacity effects on a spaxel-by-spaxel basis \citep[see ][ for further details]{Villanueva2025,Roth2023}. 

We used a fixed resolution element, $\Delta\lambda$, that was determined based on the central wavelength of each spectral extract. For 1.80 $\mu$m $\leq \lambda \leq$ 3.2 $\mu$m, we set $\Delta\lambda$ = 0.79 nm, and for 3.2 $\mu$m $< \lambda \leq $ 5.10 $\mu$m, we set $\Delta\lambda$ = 1.32 nm. These values are in good agreement with those from curves for dispersion as a function of wavelength provided by the Space Telescope Science Institute \footnote{\url{https://jwst-docs.stsci.edu/jwst-near-infrared-spectrograph/nirspec-instrumentation/nirspec-dispersers-and-filters\#gsc.tab=0}}. We assumed a gas expansion speed, $v_\mathrm{exp}$, of 0.345 \kms{} for CO and 0.310 \kms{} for all other species based on velocity resolved ALMA measurements of CO and HCN on 2025 December 22 \citep{Cordiner2026}. Uncertainties on the derived parameters were retrieved from the diagonal elements of the covariance matrix, scaled by the square root of the reduced $\chi^2$ statistic of the fit.

We generated our maps using spectra extracted from relatively unblended molecular bands of each species. This included the CO $(v=1-0)$ band centered near 4.66 $\mu$m, the \ce{CO2} $\nu_3$ band centered near 4.25 $\mu$m, the \ce{H2O} $\nu_1+\nu_3$ band near 2.68 $\mu$m, the \ce{CH3OH} $\nu_3$ band near 3.51 $\mu$m, and the \ce{CH4} $\nu_3$ band centered near 3.30 $\mu$m. Although we did not detect spectrally unblended lines of OH* \citep[prompt emission, the vibrationally excited photodissociation product of \ce{H2O} which traces the spatial distribution of the latter;][]{Bonev2006}, we conservatively included it in our models by setting $N(\ce{OH}^*)=N(\ce{H2O})$ when retrieving $N(\ce{CH3OH})$. We then fixed the assumed $N(\ce{OH}^*)$ and derived $N(\ce{CH3OH})$ at each spaxel when retrieving $N(\ce{CH4})$ and \trot{}(\ce{CH4}). We constrained our maps to within $\pm15$ spaxels of the nucleus position (corresponding to $\pm\sim1800$ km projected distance) owing to insufficient SNR at larger distances, especially for \ce{CH4} and \ce{CH3OH}.

In addition to providing the overall \ce{H2O} column density, the \ce{H2O} 2.68 $\mu$m band includes multiple strong, unblended lines of its nuclear spin isomers, ortho- and para-\ce{H2O}, enabling us to generate maps of $N(o-\ce{H2O})$ and $N(p-\ce{H2O})$. We used these to create a map of the ortho-to-para ratio (OPR) for \ce{H2O}. The NASA PSG generates $g$-factors for ortho- and para-\ce{H2O} assuming the statistical equilibrium value of 3 regardless of \trot{} \citep{Villanueva2025}. Thus, $\mathrm{OPR}= 3\times N(o-\ce{H2O}) / N(p-\ce{H2O})$ as modeled with the PSG. 

Although the \ce{H2O} hot bands near 4.5 $\mu$m measured on December 23 were detected with a sufficient signal-to-noise ratio (SNR) to derive $N$, we could not determine a well-constrained OPR or \trot{}. We instead assumed the OPR and radial dependence of \trot{}(\ce{H2O}) determined from the 2.68 $\mu$m bands on December 22 and fixed that when retrieving $N(\ce{H2O})$ on December 23. To account for several \ce{H2O} hot band lines which blend with CO, we then fixed $N(\ce{H2O})$ at each spaxel while retrieving $N(\ce{CO})$ and \trot{}(CO).

%% file: Results.tex
\section{Results} \label{sec:results}
\begin{figure}
\plotone{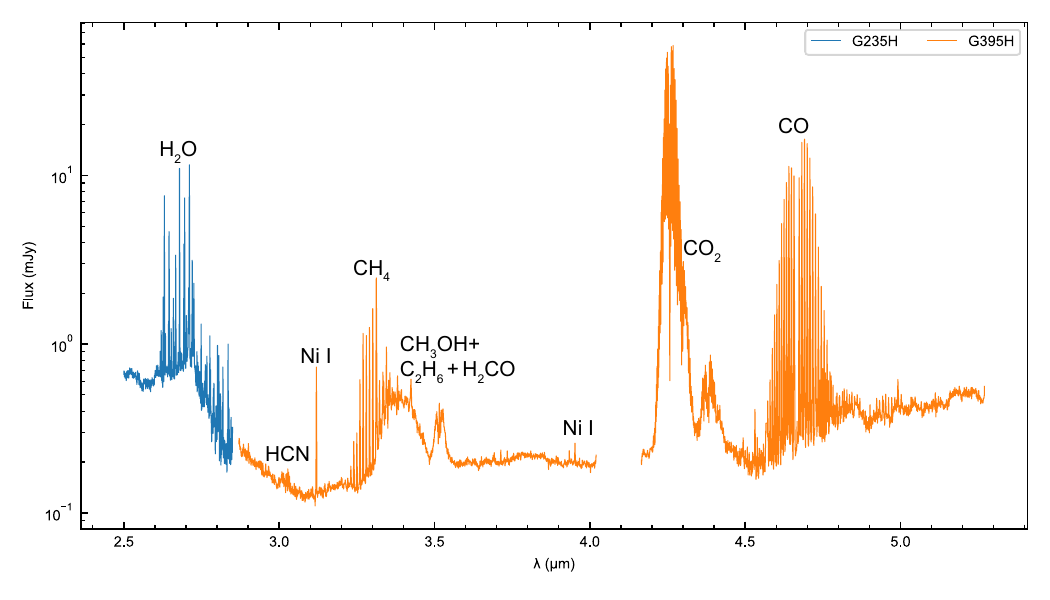}
\caption{JWST spectrum of 3I/ATLAS extracted in a $1''.5$ diameter aperture centered on the nucleus position. Major species are labeled. 
\label{fig:all}}
\end{figure}

\begin{figure}
\plotone{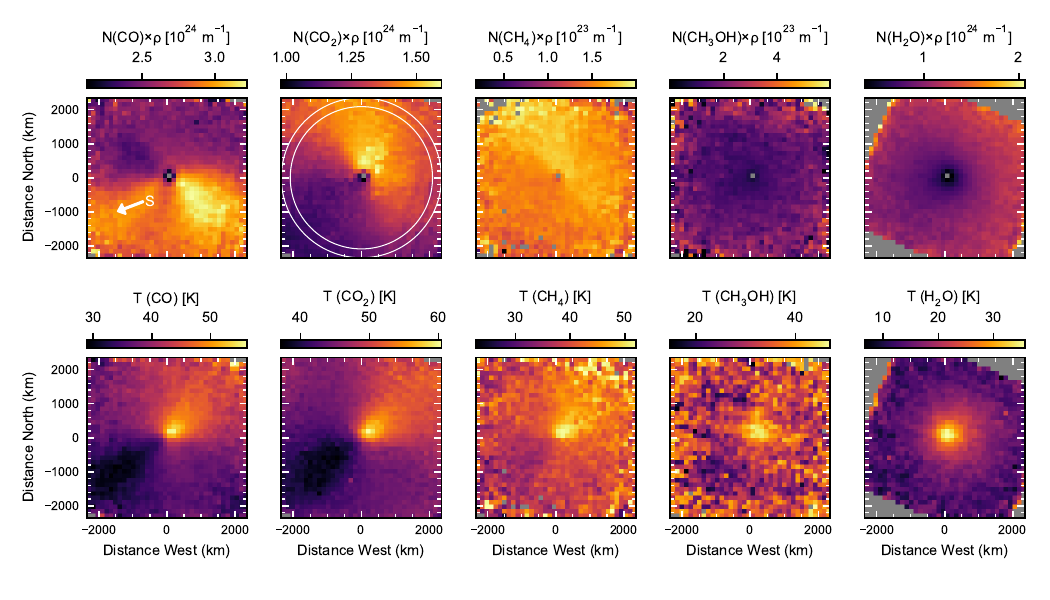}
\caption{\textbf{Upper Panels.} Maps of $N\times\rho$ for CO, \ce{CO2}, \ce{CH4}, \ce{CH3OH}, and \ce{H2O}. The white arrow shows the projected direction of the Sun. The white annulus shows the region used to calculate representative $Q$. Note that the \ce{H2O} map was derived from the 2.68 $\mu$m bands measured on December 22, whereas the remaining maps were constructed from the December 23 observations. \textbf{Lower Panels.} Maps of molecular rotational temperatures.
\label{fig:maps1}}
\end{figure}

\begin{figure}
\plotone{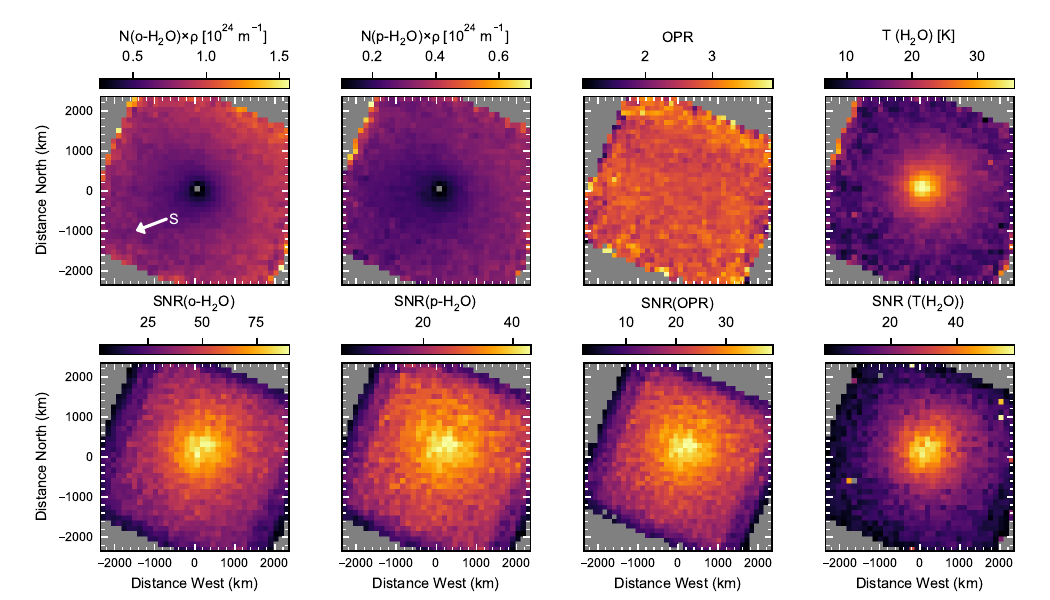}
\caption{\textbf{Upper Panels.} Maps of $N\times\rho$ for ortho-\ce{H2O} and para-\ce{H2O}, the derived OPR, and \trot{}(\ce{H2O}) for the 2.68 $\mu$  band measured on December 22. Note that $N$ for each species has been corrected for the derived OPR. \textbf{Lower Panels.} Signal-to-noise ratio (SNR) for each of the respective quantities in the upper panels.
\label{fig:maps2}}
\end{figure}

A representative JWST spectrum of 3I/ATLAS extracted from a nucleus-centered $1\farcs5$ diameter aperture is shown in Figure~\ref{fig:all}. Our resulting spaxel-by-spaxel maps for CO, \ce{CO2}, \ce{H2O}, \ce{CH3OH}, and \ce{CH4} are shown in Figure~\ref{fig:maps1}. Maps of o-\ce{H2O}, p-\ce{H2O}, and OPR are shown in Figure~\ref{fig:maps2}. The difference in spatial coverage between the 2.68 $\mu$m \ce{H2O} maps and those of the remaining species is owing to the single G235H exposure for the former vs.\ five stacked G395H exposures for the latter. A comparison of \ce{H2O} maps derived on December 22 and December 23 is given in Appendix~\ref{sec:water-maps}. Here we analyze the higher S/N \ce{H2O} emission from the 2.7 $\mu$m band measured on December 22. 

Expansion dilution of the coma leaves trends in gas distribution difficult to discern in plots of $N$. Plotting $N\times\rho$, where $\rho$ (m) is the radial distance from the nucleus, corrects for this effect and more readily reveals coma anisotropies: isotropic outgassing in the absence of photolysis and coma acceleration would produce a flat $N\times\rho$ dependence with nucleocentric distance. Differences are dramatically evident among the major constituents of 3I/ATLAS's coma as shown in Figure~\ref{fig:maps1}. Continuum maps are provided in Appendix~\ref{sec:contMaps} for comparison with the gas. A comparison of maps produced with and without the PSG correction for opacity is given in Appendix~\ref{sec:opacity}.

As $N\propto Q/v_{\mathrm{exp}}$, is important to keep in mind that asymmetries in $v_{\mathrm{exp}}$ will manifest as asymmetries in the $N\times\rho$ maps. The strong asymmetries in \trot{} for CO and \ce{CO2} mean such an asymmetry in $v_{\mathrm{exp}}$ is expected, but the JWST data do not have sufficient spectral resolution to test this. CO, the most abundant molecule by at least a factor of two, shows an apparent dual jet structure, with one oriented in the projected sunward direction and another perpendicular to it. The sunward jet is physically consistent with the lower \trot{} projected in the same direction, yet the absence of a corresponding trend in \trot{} along the perpendicular jet is puzzling. 

\ce{CO2} and \ce{CH4} display a simpler morphology with apparent excess in the projected anti-sunward direction.  The anti-sunward excess of \ce{CO2} may be an effect of an asymmetry in $v_{\mathrm{exp}}$ if the asymmetry were not sufficiently strong to be reflected in the CO map. Alternatively, the contrasting distributions of CO and \ce{CO2} may be due to the effects of rotation of differing active sites into and out of the field of view along the line of sight, as our observations covered $\sim1/3$ of 3I/ATLAS's 16.16 hr.\ rotational period \citep{Santana-Ros2025}. Analysis of time-series spectra is beyond the scope of this manuscript and the subject of a future publication. 

In contrast, \ce{H2O} and \ce{CH3OH} (although at lower SNR for the latter) show $N\times\rho$ maps which are much more symmetric. Although there is an appearance of increasing $N\times\rho$ with $\rho$ for these species, which may be consistent with production from extended sources such as icy grain sublimation, it is important to note that the innermost spaxels are affected by significant flux losses due to the finite PSF, and the significance of the trend is unclear. However, \cite{Cordiner2026} found evidence for an \ce{H2O} extended source based on ``Q-curve'' analysis of the less optically thick 4.5 $\mu$m hot bands.

The spatial variations in \trot{} are similarly striking and vary from molecule to molecule. CO, \ce{CO2}, and \ce{CH4} show a common temperature morphology, having distinctly higher values in the projected anti-sunward direction. On the whole, these molecules have higher \trot{} (and maintain it to greater nucleocentric distances) than \ce{CH3OH} or \ce{H2O}. In contrast, the \ce{H2O} temperature peaks at the nucleus position and drops quickly and nearly isotropically. \ce{CH3OH} appears to follow a similar spatial morphology to \ce{H2O}, but SNR limitations preclude a detailed inter-comparison between these species.

An average column density, $\langle N \rangle$, was calculated by averaging individual column densities derived for each spaxel within a $0\farcs2$ wide annulus centered on the nucleus, which was defined as the continuum photocenter, with $r_{\mathrm{in}}=1\farcs6$ and $r_{\mathrm{out}}=1\farcs8$. These $\langle N \rangle$ were converted to representative molecular production rates ($Q$, \ps{}) following \cite{Bonev2017} as:

\begin{equation}\label{eqn:eqn1}
Q = \frac{\langle N \rangle A_{\mathrm{FOV}}}{\tau(r_\mathrm{H}) f(x)}
\end{equation}

where $A_{\mathrm{FOV}}$ is the area of the aperture (m$^2$), $\tau(r_{\mathrm{H}})$ is the molecular photolysis lifetime at \rh{} \citep{Huebner2015}, and $f(x)$ is the Haser filling factor (the fraction of coma molecules expected within the aperture after correcting for adiabatic expansion and solar photolysis). The resulting production rates are given in Table~\ref{tab:qs}.

\begin{deluxetable*}{ccccc}
\tablenum{1}
\tablecaption{\label{tab:qs} Molecular Production Rates in 3I/ATLAS and Comets Measured}
\tablewidth{0pt}
\tablehead{
\colhead{Molecule} & \colhead{$Q$} & \colhead{$Q_x/Q$(CO)} & \colhead{$Q_x/Q(\ce{H2O})$} & \colhead{$\langle Q_x/Q(\ce{H2O})\rangle$}  \\
\colhead{} & \colhead{($10^{26}$ \ps{})} & \colhead{(\%)} & \colhead{(\%)} & \colhead{(\%)} 
}
\startdata
\multicolumn{5}{c}{2025 December 22, G235H/F170LP} \\
\ce{H2O} & $15.7\pm0.1$ & ... & 100 & 100 \\
\hline
\multicolumn{5}{c}{2025 December 23, G395H/F290LP} \\
CO & $38.2\pm0.1$ &  100 & $246\pm19$ & $5.2\pm1.3$  \\
\ce{CO2} & $15.9\pm0.1$ & $41.6\pm0.3$ & $103\pm8$ & $12\pm2$  \\
\ce{H2O} & $15.5\pm1.2$ & $40.6\pm3.1$ & 100 & 100   \\
\ce{CH3OH} & $2.65\pm0.03$ & $6.94\pm0.09$ & $17.1\pm1.3$ & $2.06\pm0.20$   \\
\ce{CH4} & $2.06\pm0.01$ & $5.39\pm0.03$ &  $13.3\pm1.0$ & $0.78\pm0.09$   \\
\enddata
\tablecomments{$\langle Q_x/Q(\ce{H2O})\rangle$ are average molecular abundances with respect to water in comets reported in \cite{Harrington2022,DelloRusso2016}.}
\end{deluxetable*}

Our map of the OPR for \ce{H2O} demonstrates an essentially flat value with nucleocentric distance, independent of changes in the total column density, \trot{}, or SNR. We derived a coma-averaged value of $\mathrm{OPR}=2.7\pm0.2$ by calculating statistics for values retrieved within a 10-spaxel radius of the comet photocenter (Appendix~\ref{sec:modeling}). 

%% file: Discussion.tex
\section{Discussion and Interpretation} \label{sec:discussion}
\subsection{Evolving Heterogeneous Outgassing}\label{subsec:evolving}
Our measurements reveal a CO-driven coma in 3I/ATLAS ($\ce{CO2}/\ce{CO}=(42.4\pm0.9)\%$ and $\ce{H2O}/\ce{CO}=(44.4\pm0.7)\%$) with a complex outgassing geometry which varies from species to species. The volatile composition of 3I/ATLAS during our observations is in general super-enriched compared to water with respect to values found in solar system comets \citep{DelloRusso2016,Harrington2022,Biver2024b}. Our measurements provide a rare characterization of the interplay between $N$ and \trot{} for a comet at \rh{}$\sim2.4$ au post-perihelion, and thus traversing the \ce{H2O} ice line. 

At these distances, \ce{H2O} sublimation becomes less vigorous as solar insolation attenuates, leaving CO and \ce{CO2} to gradually overtake \ce{H2O} as the primary activity drivers. The analysis of these three species in \cite{Cordiner2026} is consistent with 3I/ATLAS's transition from \ce{H2O}- to CO-dominated outgassing as it receded from the Sun. Since the majority of comet composition studies are undertaken interior to \rh{}$\sim3$ au in \ce{H2O}-dominated comae \citep[e.g.,][ and references therein]{DelloRusso2016,Lippi2021}, this caveat must be kept in mind, as $Q(x)/Q(\ce{H2O})$ in 3I/ATLAS during our observations may no longer be strictly indicative of bulk nucleus ice abundances. On the other hand, serial pre-perihelion ALMA measurements and near-perihelion IRAM 30-m measurements of \ce{CH3OH} found that it was significantly enriched compared to solar system comets \citep{Roth2026a,Biver2026}, consistent with the interpretation that 3I/ATLAS was generally enriched in trace volatiles regardless of activity driver. 

Our representative $Q(\ce{CO})$ and $Q(\ce{CO2})$ are each slightly lower than reported in \cite{Cordiner2026}, whereas our $Q(\ce{H2O})$ on December 22 is higher, but the value on December 23 agrees with theirs within $1\sigma$. Both studies are consistent with a CO-dominated coma where \ce{H2O}/CO and \ce{CO2}/CO were on the order of $40\%$. These differences may be explained by differing methodologies for each study: whereas \cite{Cordiner2026} constructed a ``Q-curve'' and calculated terminal production rates based on the averages of several annuli, we report $Q$ calculated from a single outer annulus. Similarly, our $Q(\ce{CH4})$ and $Q(\ce{CO2})$ are lower than reported in \cite{Belyakov2026} based on JWST MIRI observations taken five days before and four days after our observations, although they hypothesized that $Q(\ce{CH4})$ was quite variable during mid-late December. The differences in $Q(\ce{CO2})$ may be attributed to additional uncertainties in modeling the 15 $\mu$m \ce{CO2} band measured with MIRI, which is superimposed on strong thermal emission, compared to the 4.25 $\mu$m band measured with NIRSpec in this study. Future JWST observations of cometary \ce{CO2} with significantly smaller temporal separation between MIRI and NIRSpec measurements may help to reconcile potential uncertainties when modeling the emission bands sampled by each instrument.

The molecule-dependent coma morphology observed in 3I/ATLAS resembles the heterogeneous jet structures identified in the Centaur 29P/Schwassmann–Wachmann 1 \citep{Faggi2024}. However, unlike 29P, whose activity is driven primarily by hypervolatiles at a heliocentric distance of $\approx$5~au, the JWST observations of 3I/ATLAS were obtained following intense perihelion activity, during which water production reached $\sim 2 \times 10^{29}$ \ps{} \citep{Combi2026}. The observed volatile segregation may therefore reflect the exposure of chemically heterogeneous subsurface reservoirs following substantial perihelion mass loss rather than purely primordial compositional heterogeneity, similar to the evolution of comets C/2009 P1 (Garradd) and 2I/Borisov \citep{Bodewits2014, Bodewits2020}.

\input{Physics}
\input{ortho-para}

%% file: Physics.tex
\subsection{Coma Physics of an Interstellar Object}\label{subsec:physics}
The coma physics in 3I/ATLAS revealed by our maps are just as remarkable as its molecular abundances. It is important to note that these maps are inherently biased to emission perpendicular to the line of sight owing to the low spectral resolution of JWST. Velocity-resolved measurements from radio facilities \citep[e.g.,][]{Biver2026,Cordiner2026} are required to incorporate a detailed treatment of emission along the line of sight. Such an integration and analysis is the subject of a future publication. Nevertheless, the CO, \ce{CO2}, and \ce{CH4} rotational temperatures show a clear dichotomy along an axis similar to the sky-projected Sun-comet radius vector, with the anti-sunward direction providing considerably higher values. Similar behavior has been observed in other comets, namely joint JWST and ALMA studies of comet C/2022 E3 \citep[ZTF;][]{Foster2026} and ALMA and IRAM 30-m studies of comet 46P/Wirtanen \citep{Cordiner2023,Biver2021}. This was explained by a combination of more efficient adiabatic cooling in the sunward-facing hemisphere of the coma, along with the heating effects of icy grain sublimation in the anti-sunward direction. 

Most comets are characterized via remote sensing when they are within $\sim3$ au from the Sun \citep[e.g.,][]{Lippi2021,DelloRusso2016}. For these studies, \ce{H2O} is the most abundant coma molecule and dominates excitation in the inner coma \citep[$r<1000$ km;][]{Tenishev2008,Fougere2012,Marschall2024}, where collisions thermalize the rotational temperature to the kinetic temperature of the gas. Yet \ce{H2O} was a trace species during our observations of 3I/ATLAS, and the spatial distributions of its column density and rotational temperature stand apart from all other molecules (except \ce{CH3OH}) in this study. These differences in coma temperature distributions can be understood in terms of the overall abundances and radiative cooling efficiencies of each species. CO is the most abundant molecule and plays an outsized role in setting the temperature structure of the coma. However, \ce{CO2} and \ce{H2O} are both $\sim40\%$ of CO and must contribute as well. The apolar species are likely the best tracers of the kinetic temperature profile of the inner, collisional coma given their small dipole moments (and thus lower radiative cooling efficiency), whereas the rotational temperature profiles for the polar species \ce{H2O} and \ce{CH3OH} depart from the kinetic temperature more quickly \citep{Bodewits2024}. 

\begin{figure}
\plotone{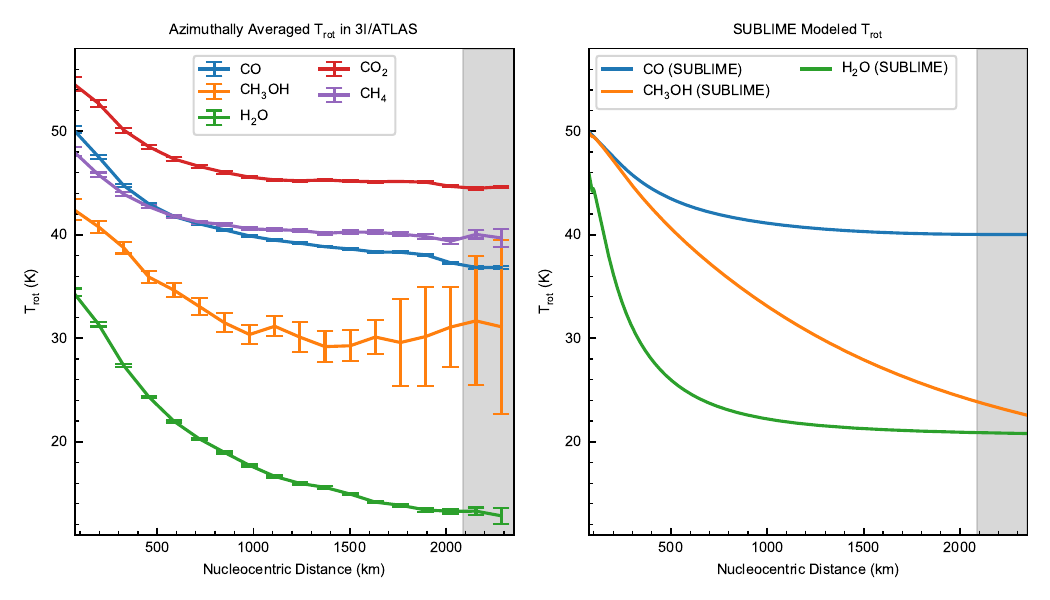}
\caption{\textbf{Left.} Observed azimuthally averaged \trot{} for CO, \ce{CO2}, \ce{CH4}, \ce{CH3OH}, and \ce{H2O}. \textbf{Right.} Modeled \trot{} as a function of nucleocentric distance for CO, \ce{CH3OH}, and \ce{H2O} using the SUBLIME radiative transfer code. The gray shaded region shows the projected radial extent of the annulus from which representative $Q$'s were extracted for each molecule (Table~\ref{tab:qs}). The $x$-axis extends to approximately the field of view of our spatial-spectral maps.
\label{fig:temps}}
\end{figure}

We calculated azimuthally averaged radial \trot{} profiles for CO, \ce{CH3OH}, and \ce{H2O} in 3I/ATLAS and compared them against predictions from the three-dimensional SUBLIME radiative transfer and excitation model for cometary atmospheres \citep{Cordiner2022}. Our methods for generating the SUBLIME models are detailed in Appendix~\ref{sec:sublime}. It is important to note that our spherically symmetric SUBLIME models do not attempt to reproduce the complex, anisotropic, and species-dependent outgassing geometry evident in Figure~\ref{fig:maps1} and are only intended to test whether the general trends observed in the coma of 3I/ATLAS can be reproduced with state-of-the-art radiative transfer models. Figure~\ref{fig:temps} shows a progression for each species that indeed qualitatively matches our observations. The CO model drops by only $\sim8$ K within 1750 km from the nucleus, yet remains $\sim2$ K warmer than the observations. In contrast, the \ce{H2O} model falls out of collisional equilibrium quickly and dramatically owing to its higher dipole moment and radiative cooling efficiency, whereas \ce{CH3OH} follows an intermediate trend in cooling. The observed \ce{CH3OH} \trot{} cools significantly more quickly than the models close to the nucleus. Similarly, the observed \ce{H2O} \trot{} profile is cooler than the model close to the nucleus, yet the model's steep initial cooling levels off at $\sim1000$ km from the nucleus whereas the observed profile continues to drop. The slower drop in the observed temperature compared to the model may be a signature of sublimative gas heating from icy grains in the coma, consistent with the results of \cite{Cordiner2026}. Additional cooling through mechanisms such as coma gas acceleration may explain these trends, but these effects are not considered in our models. These observational and model results in 3I/ATLAS for polar vs.\ apolar species as a function of increasing nucleocentric distance are consistent with predictions for comets in general \citep{Bodewits2024}.

The rotational temperature distributions are remarkably similar to results found by JWST for comet C/2017 K2 (PanSTARRS), where the temperature distributions of \ce{H2O} and \ce{CH3OH} were closely matched, yet \ce{CH4} and (to a lesser extent) CO followed that of \ce{CO2}, despite \ce{CO} and \ce{CH4} having considerably different column density distributions from \ce{CO2} \citep{Woodward2025}.  The trends in the overall magnitude of the CO, \ce{CO2}, and \ce{CH4} temperatures compared to \ce{H2O} and \ce{CH3OH} in 3I/ATLAS were also very similar to C/2017 K2, with the former apolar species being higher (and staying so) out to considerably larger nucleocentric distances than the latter polar species. This was explained in terms of the increased rotational cooling efficiency for polar \ce{H2O} and \ce{CH3OH} compared to apolar CO, \ce{CO2}, and \ce{CH4} \citep{Woodward2025}. It is interesting to note that observations of C/2017 K2 were conducted when \ce{H2O} was the dominant volatile, whereas our observations of 3I/ATLAS were of a CO-driven coma.

%% file: ortho-para.tex
\subsection{Ortho-to-Para Ratio and Spin Temperatures in Comets}\label{subsec:ortho-discuss}
Despite its relegation to a trace species during our observations of 3I/ATLAS, \ce{H2O} provides insights into the nature of the OPR in comets. Water, like many hydrogen-bearing molecules, exists in distinct nuclear spin symmetry forms known as ortho and para species. These forms arise from the fundamental quantum mechanical requirement that the total molecular wavefunction obey specific symmetry constraints under exchange of identical nuclei. In \ce{H2O}, the two equivalent protons give rise to ortho and para states with statistical weights in a 3:1 ratio. Under thermodynamic equilibrium, the OPR depends on temperature. At high temperatures, the OPR approaches the statistical value of three, whereas at low temperatures the population shifts toward the lowest-energy symmetry species. The temperature that reproduces a measured OPR under equilibrium conditions is referred to as the spin temperature. 

A key property of ortho and para species is their radiative independence. Electric dipole transitions conserve nuclear spin symmetry; consequently, ortho levels are connected only to other ortho levels, and para levels only to para levels. Radiative processes therefore cannot inter-convert the two species. Conversion requires specific non-radiative mechanisms, such as proton exchange reactions, interactions with paramagnetic species, catalytic processes on surfaces, or magnetic interactions. In low-density astrophysical environments and cometary comae, these processes may proceed slowly, allowing the OPR to become effectively ``frozen''.

This radiative isolation historically motivated the interpretation that the measured spin temperature may reflect the temperature at which the molecule formed or last achieved nuclear spin equilibrium \citep[e.g.,][]{Mumma1987,Bonev2007,BockeleeMorvan2009,Villanueva2011a}. However, recent laboratory studies suggest a more complex picture, indicating that the OPR may be reset upon desorption \citep[e.g.,][]{Hama2018} and that spin symmetry properties can be partially altered during phase transitions between ice and gas \citep[e.g.,][]{Yocum2023}.

In the case of 3I/ATLAS, we find that the OPR remains equilibrated and notably constant throughout the coma, even as the rotational temperature falls below 30 K. This represents a significant new observational result in the study of cometary OPRs. Although previous works \citep{Bonev2007,BockeleeMorvan2009} have found similar results at higher temperatures, none have demonstrated these trends at such cold temperatures. The constancy of the ratio indicates that it is not being reset within the coma and instead reflects conditions associated with the release of water from the nucleus. The coma-averaged $\mathrm{OPR}=2.7\pm0.2$ corresponds to $T_{\mathrm{spin}}=34^{+11}_{-5}$ K, yet whether the measured OPR directly traces the formation conditions or origin of the object remains uncertain. Nevertheless, the 3I/ATLAS observations show that rapid physical processes in the expanding coma, including substantial adiabatic cooling, are insufficient to alter the statistical spin symmetry distribution of the molecule once it has been established.

%% file: Conclusion.tex
\section{Conclusion} \label{sec:conclusion}
The sensitivity and spatial-spectral mapping capabilities of JWST have revealed the coma physics of an interstellar object as it receded from its encounter with our Sun. Our spaxel-by-spaxel maps show stark dichotomies in the column density and rotational temperature distributions of the apolar vs.\ polar species in 3I/ATLAS' coma. CO was the most abundant molecule and set the overall temperature structure of the coma, yet the increased radiative cooling efficiency of the polar molecules caused them to fall out of thermal equilibrium much more quickly than their apolar counterparts, resulting in significantly different \trot{} distributions for \ce{CH3OH} and \ce{H2O} compared to CO, \ce{CO2}, and \ce{CH4}. Our high-resolution mapping of the OPR in \ce{H2O} measured a flat dependence with nucleocentric distance even at low ($<20$ K) temperatures, confirming that the OPR is not altered in the coma. Reproducing the complexity of coma excitation revealed in 3I/ATLAS will require new detailed kinetic modeling as JWST measurements of \ce{H2O} and \ce{CO2} in comets challenge our understanding of coma physics.

%% file: obsAppendix.tex
\section{Observing Log}\label{sec:obslog}
Table~\ref{tab:obslog} provides an observing log for our JWST observations of 3I/ATLAS.

\begin{deluxetable*}{ccccccc}
\tablenum{2}
\tablecaption{Observing Log\label{tab:obslog}}
\tablewidth{0pt}
\tablehead{
\colhead{Date} & \colhead{UT Time} & \colhead{\textit{T}\subs{int}} &
\colhead{\textit{r}\subs{H}} & \colhead{$\Delta_{\mathrm{JWST}}$} & \colhead{$\phi_\mathrm{STO}$} & \colhead{$\lambda$} \\
\colhead{(2025)} & \colhead{} & \colhead{(s)} & \colhead{(au)} & 
\colhead{(au)} & \colhead{($\degr$)} & \colhead{($\mu$m)}
}
\startdata
22 Dec & 03:36 & 642 & 2.37 & 1.79 & 22.7 & $1.06-3.05$ \\
23 Dec & 08:07 & 700 & 2.40 & 1.80 & 21.8 & $2.87-5.14$ \\
       & 09:57 & 700 & 2.41 & 1.80 & 21.8 & $2.87-5.14$ \\
       & 11:43 & 700 & 2.41 & 1.80 & 21.7 & $2.87-5.14$ \\
       & 12:56 & 700 & 2.41 & 1.80 & 21.7 & $2.87-5.14$ \\
       & 13:58 & 700 & 2.41 & 1.80 & 21.7 & $2.87-5.14$ \\
\enddata
\tablecomments{UT times are given at the start of each exposure. \textit{T}\subs{int} is the total on-source integration time. \textit{r}\subs{H}, $\Delta_{\mathrm{JWST}}$, and $\phi_\mathrm{STO}$, and $\psi_{\sun}$ are the heliocentric distance, JWST-centric distance,
and phase angle (Sun--Comet--Earth), respectively, of 3I/ATLAS at the time of observations. }
\end{deluxetable*}

%% file: waterAppendix.tex
\section{Comparison of \ce{H2O} Maps on December 22 and December 23}\label{sec:water-maps}

\begin{figure}
\plotone{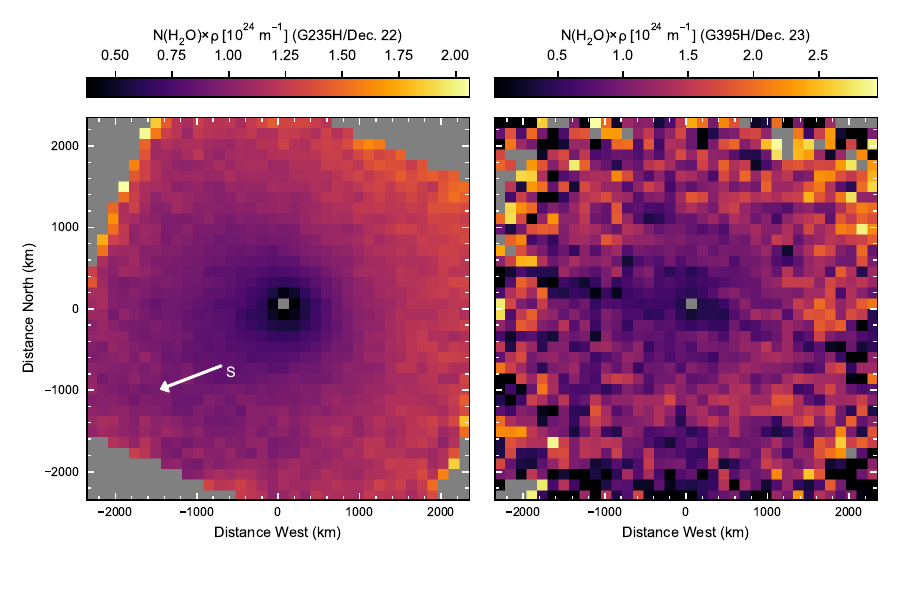}
\caption{\textbf{Left.} \ce{H2O} map in 3I/ATLAS on December 22. \textbf{Right.} \ce{H2O} map in 3I/ATLAS on December 23.
\label{fig:water-diff}}
\end{figure}

Figure~\ref{fig:water-diff} shows a comparison of \ce{H2O} maps in 3I/ATLAS on December 22 (2.7 $\mu$m band measured with G235H) and December 23 (4.5 $\mu$m band measured with G395H).

%% file: contAppendix.tex
\section{Continuum Maps in 3I/ATLAS}\label{sec:contMaps}

\begin{figure}
\plotone{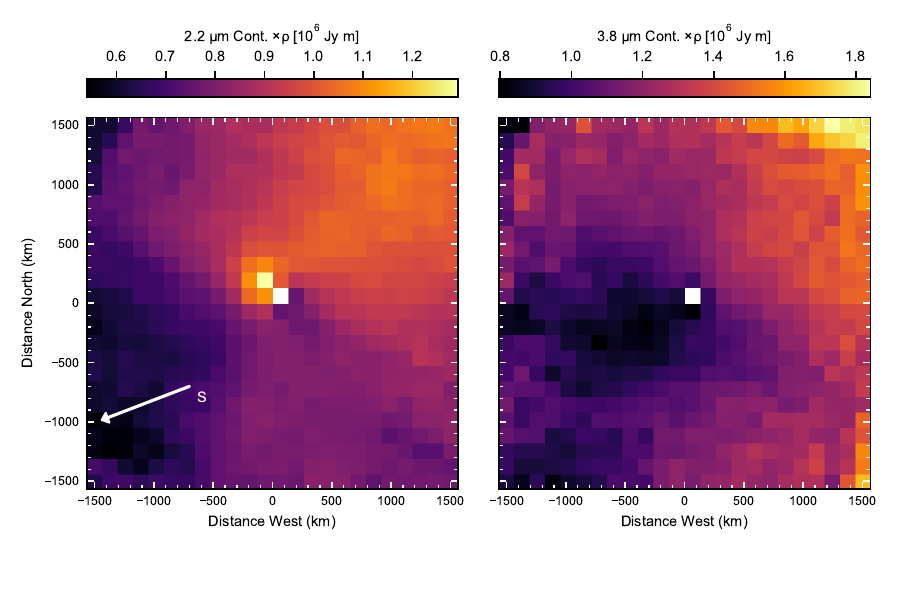}
\caption{\textbf{Left.} Map of the 2.2 $\mu$m continuum intensity, measured on December 22, multiplied by $\rho$. \textbf{Right.} Map of the 3.8 $\mu$m continuum intensity, measured on December 23, multiplied by $\rho$.
\label{fig:contMaps}}
\end{figure}

Figure~\ref{fig:contMaps} shows continuum maps in 3I/ATLAS (multiple by $\rho$) for the G235H (December 22) and G395H (December 23) settings.

%% file: opacityAppendix.tex
\section{Opacity Corrections to Column Density in the PSG}
\label{sec:opacity}
We refer to \cite{Villanueva2025,Roth2023} for a detailed discussion of the opacity corrections performed by the PSG. Here we give a brief overview of the formalism and document how the corrections affected our calculated $N$ for the major species in 3I/ATLAS' coma during our observations: CO, \ce{CO2}, and \ce{H2O}.

For a low solar phase angle, the integrated column density measured by an observer is approximately the column density of the incident solar flux.  The average linewidth of the pump ($w_p$, cm$^{-1}$) can be computed as $w_p$ = (2$v_pf_{ul}/c$), where $v_p$ is the gas expansion velocity (m s$^{-1}$), $f_{ul}$ is the emission frequency (cm$^{-1}$), and $c$ is the speed of light. The integrated opacity across the pump is then $\tau_p=N_{\mathrm{col}}S_p/v_p$, where $N_{\mathrm{col}}$ is the integrated column density (molecules cm$^{-2}$) along the line of sight and $S_p$ is a weighted ``representative'' line intensity (cm$^{-1}$ molecule cm$^{-2}$). This weighted line intensity is calculated after keeping track of the associated line intensities that result in the specific emission for every $g$-factor (photon s$^{-1}$ molecule$^{-1}$) and calculating a corresponding weighted pump intensity.

\begin{figure}
\plotone{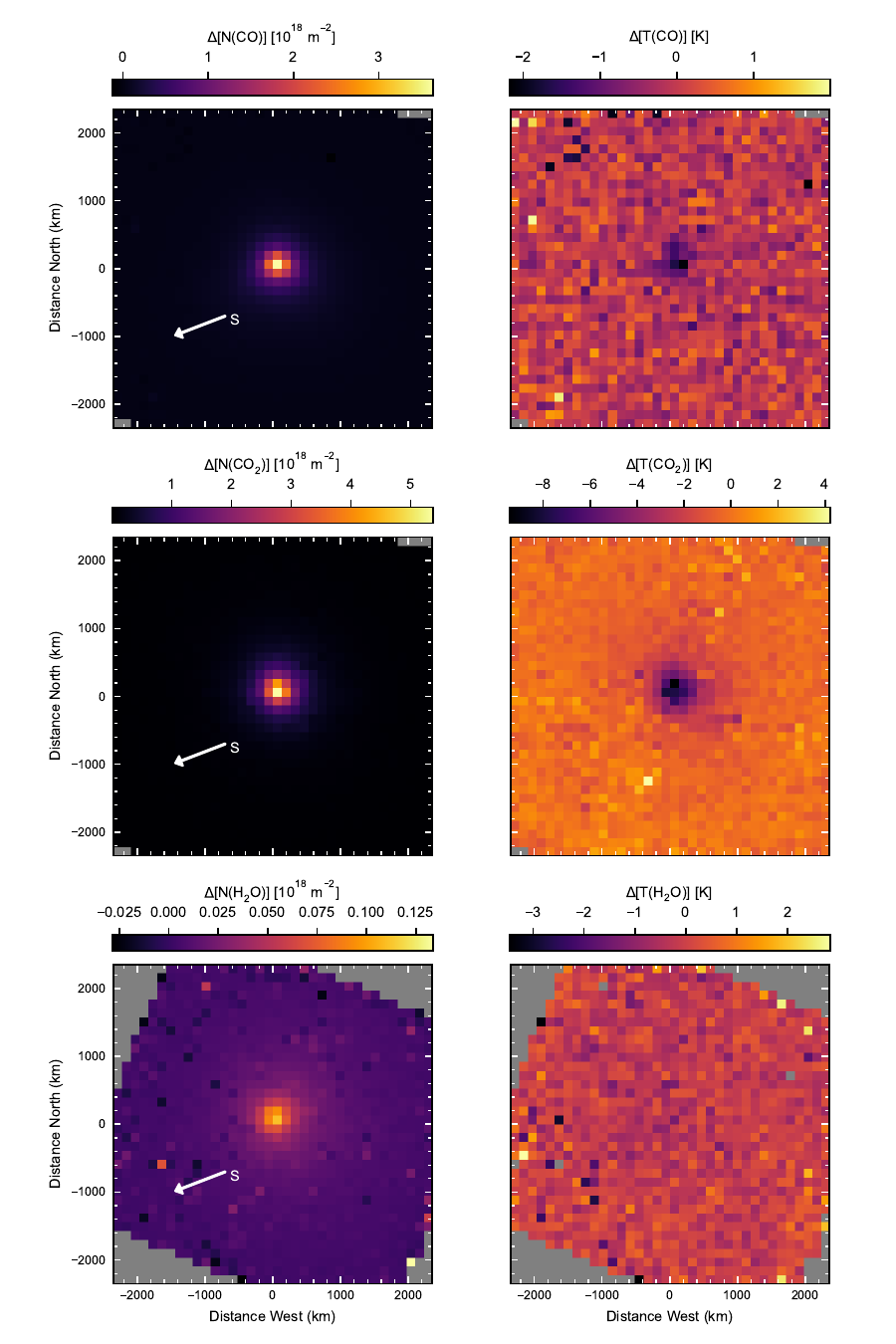}
\caption{\textbf{From top.} Maps of differences between $N$ and \trot{} for optically thick and optically thin PSG retrievals for CO, \ce{CO2}, and \ce{H2O}, where $\Delta N=N_{\mathrm{thick}}-N_{\mathrm{thin}}$ and $\Delta T=T_{\mathrm{thick}}-T_{\mathrm{thin}}$.
\label{fig:opacity}}
\end{figure}

The transmittance at the end of the column is $e^{-\tau_p}$ at the frequency of the pump. For low opacities, the pumps are optically thin given the small number of molecules attenuating the solar pump at the pump frequency. As the opacity increases along the column for the solar pump, the observer only receives radiation up to $\tau_p<1$, so the expected fluorescence efficiency can be approximated as

\begin{equation}
g_{\mathrm{thick}}=\frac{1-e^{-\tau_p}}{\tau_p}g_{\mathrm{thin}}
\end{equation}

We note that this is a first-order correction to a complex problem. Applying the correction, the optically thin column density is related to the optically thin column density and optically thin $g$-factors ($g_i$) as

\begin{equation}
N_\mathrm{thick} = \frac{\Sigma_i \frac{(1-e^{-\tau_i})}{\tau_i} g_i}{\Sigma_i g_i} N_{\mathrm{thin}}
\end{equation}

We applied these corrections to all retrievals in this study when generating our maps. We also generated maps forcing the use of optically thin $g-$factors for comparison. Figure~\ref{fig:opacity} shows the differences in retrieved quantities for our maps for CO, \ce{CO2}, and \ce{H2O}. As expected, the most significant differences occur in the central spaxels (where PSF losses are also at play) and for CO and \ce{CO2}. Differences for \ce{H2O} are much smaller, consistent with its lower opacity throughout the NIRSpec field of view. We note that the annulus used to calculate representative $Q$'s is located in the outer regions of the field of view, where agreement between the optically thick and optically thin quantities is good for all species.

%% file: ModelingAppendix.tex
\section{Coma Averaged \ce{H2O} OPR}\label{sec:modeling}
\begin{figure}
\plotone{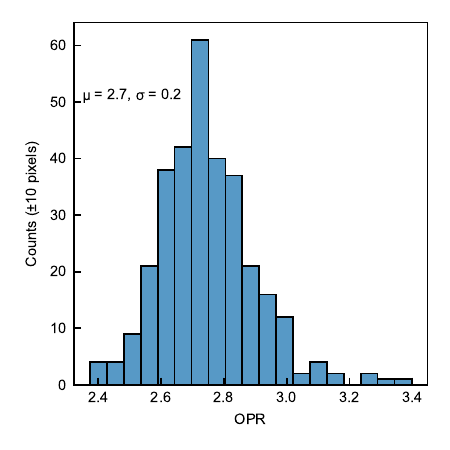}
\caption{Histogram of spaxel-by-spaxel OPR values for \ce{H2O} drawn within a 10-spaxel radius of the comet photocenter.
\label{fig:opr}}
\end{figure}

Figure~\ref{fig:opr} shows our coma-averaged OPR generated by considering statistics within a 10-spaxel radius of the photocenter.

%% file: sublimeAppendix.tex
\section{SUBLIME Radiative Transfer Models}\label{sec:sublime}
The SUBLIME radiative transfer model includes a full non-LTE treatment of coma gases, collisions with molecules and electrons, and pumping by solar radiation, along with a time-dependent integration of the energy level population equations. We calculated collisional rates for each species with CO (the dominant coma molecule during our observations) using the thermalization approximation \citep{Crovisier1987,Biver1999,Bockelee-Morvan2012}, using an average collisional cross-section with CO of $5\times10^{-14}$ cm$^{-2}$ and an electron collisional scaling factor $x_{\mathrm{ne}}=0.2$ \citep{Cordiner2025a}. We used a spherically symmetric outgassing model and set the gas expansion velocity to constant values of 0.345 \kms{} for CO and 0.310 \kms{} for \ce{CH3OH} and \ce{H2O} \citep{Cordiner2026}. We set molecular production rates for each species to their values in Table~\ref{tab:qs}. We generated a radial kinetic temperature profile matching the observed azimuthally averaged \trot{} profile for CO. We then calculated the model \trot{} as a function of radius for each species using molecular linelists from the LAMDA database \citep{Schoier2005,Rabli2010}. We included rotational energy levels up to $E_u\leq4500$ K for CO, $E_u\leq1400$ K for \ce{H2O}, and $E_u\leq400$ K for \ce{CH3OH}. These cutoffs minimize computational complexity while still adequately sampling the rotational energy manifold at the $\sim20-50$ K rotational temperatures in the coma of 3I/ATLAS. 

%% file: hcnAppendix.tex
\section{HCN and Ni in 3I/ATLAS}\label{sec:hcn}
\begin{figure}
\plotone{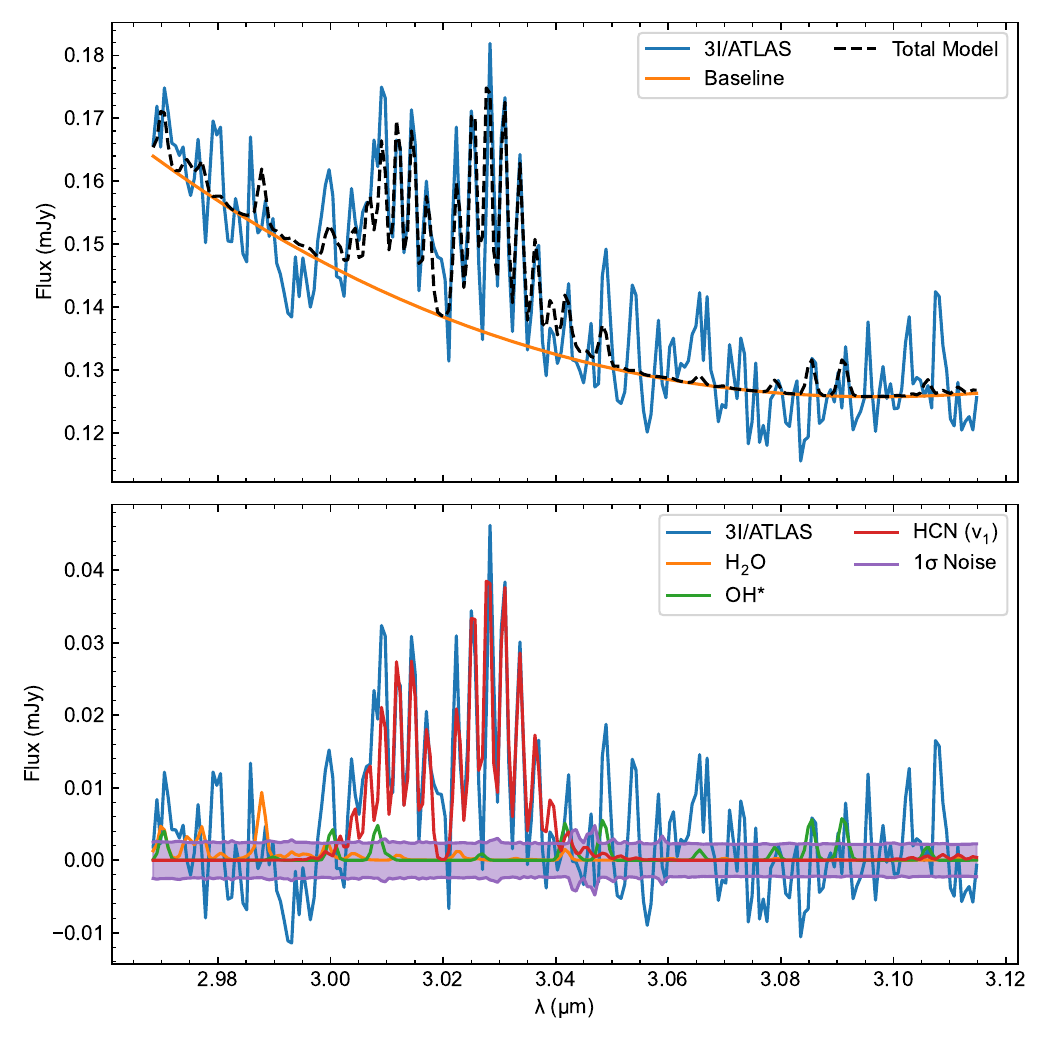}
\caption{\textbf{Upper.} 3I/ATLAS 3 $\mu$m spectrum with total molecular emission model and spectral baseline shown. \textbf{Lower.} Baseline-subtracted spectrum with individual molecular models and the $1\sigma$ noise envelope overplotted. 
\label{fig:hcn}}
\end{figure}

We additionally detected HCN and Ni I in 3I/ATLAS; however, the SNR of the former at off-nucleus positions was insufficient for spaxel-by-spaxel mapping. We instead extracted a single spectrum from a $1\farcs5$ diameter nucleus-centered aperture (Figure~\ref{fig:hcn}) and retrieved $Q(\ce{HCN})$ and \trot{}(HCN) with the PSG. We obtain $Q(\ce{HCN})=(8.22\pm0.67)\times10^{24}$ \ps{} and \trot{}(HCN) = $27\pm3$ K. The resulting molecular abundance ratios are HCN/CO = $(0.22\pm0.02)\%$ and HCN/\ce{H2O}=$(0.53\pm0.06)\%$. This \trot{} for polar HCN is consistent with the trends measured for \ce{CH3OH} and \ce{H2O} compared to the apolar species.

\begin{deluxetable*}{ccccccc}
\tablenum{3}
\tablecaption{Detected Ni I Lines in 3I/ATLAS\label{tab:ni}}
\tablewidth{0pt}
\tablehead{
\colhead{$\lambda$} & $\int F_\lambda d\lambda$ & $g$ & $A$ & Lower Level & Upper Level & Confidence \\
\colhead{($\mu$m)} & \colhead{($10^{-19}$ W m$^{-2}$)} & \colhead{($10^{-24}$ J s$^{-1}$ mol$^{-1}$)} & \colhead{($10^{-3}$ s$^{-1}$)} & \colhead{} & \colhead{} & \colhead{}
}
\startdata
3.119 & $17.3\pm0.1$ & $112\pm34$ & 78 & $3d^9(^2D)4s(^3D_3)$ & $3d^9(^2D)4s(^1D_2)$ & D \\
3.915 & $1.21\pm0.06$ & $7.0\pm2.1$ & 6.2 & $3d^9(^2D)4s(^3D_2)$ & $3d^9(^2D)4s(^1D_2)$ & D \\
\enddata
\tablecomments{$\lambda$ is the lab-measured wavelength of each transition. $\int F_\lambda d\lambda$ is the integrated intensity of each transition in 3I/ATLAS extracted from a nucleus-centered $1\farcs5$ diameter aperture. $g$ is the calculated g-factor and $A$ is the Einstein-A coefficient. The lower and upper state configurations for each transition are given. Einstein-A coefficients for both lines are rated with ``D'' confidence level in the NIST ASD \citep{Kramida2024} indicating 50\% confidence in the calculated coefficient.}
\end{deluxetable*}

\begin{figure}
\plotone{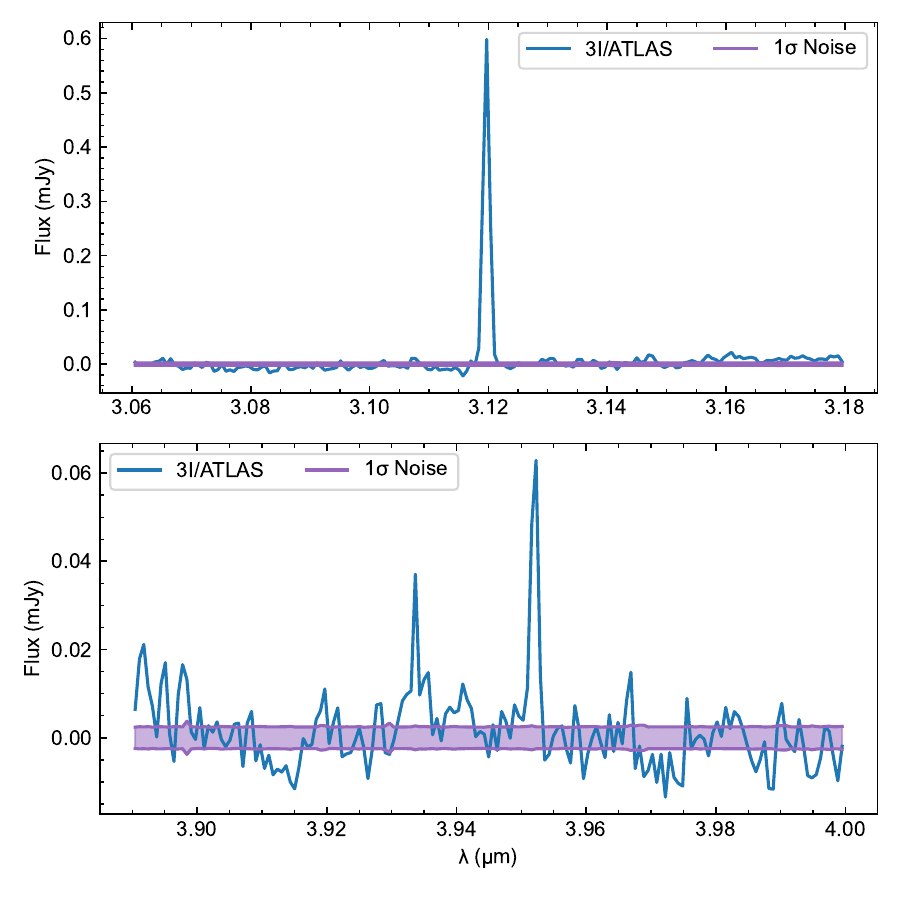}
\caption{Baseline-subtracted Ni I 3.119 $\mu$m and 3.951 $\mu$m spectra in 3I/ATLAS extracted from a nucleus-centered $1\farcs5$ diameter aperture. 
\label{fig:ni}}
\end{figure}

For Ni I, we detected the 3.119 $\mu$m and 3.915 $\mu$m  transitions, with the former sufficiently bright for spaxel-by-spaxel mapping. 3I/ATLAS has been noted as an Ni- and Fe-rich object based on studies at optical wavelengths \citep{Hutsemekers2026}. We first analyzed spectra of both transitions extracted from a nucleus-centered $1\farcs5$ diameter aperture (Figure~\ref{fig:ni}). We calculated $g-$factors and $\langle N \rangle$ for each transition following \cite{Bromley2021}, then calculated $Q(\ce{Ni})$ using Equation~\ref{eqn:eqn1}. We obtained $\langle N \rangle=(3.66\pm0.02)\times10^{15}$ m$^{-2}$ and $\langle N \rangle=(4.1\pm0.2)\times10^{15}$ m$^{-2}$ for the 3.119 $\mu$m and 3.951 $\mu$m transitions, respectively. We calculated $f(x)$ using active sun photodissociation rates for Ni \citep{Huebner2015} along with our assumed $v_{exp}=0.310$ \kms{} and assuming a range of parent scale lengths from 0 km to 2000 km, giving values ranging from $Q(\ce{Ni})=(2.2-6.8)\times10^{24}$ \ps{}.

\begin{figure}
\plotone{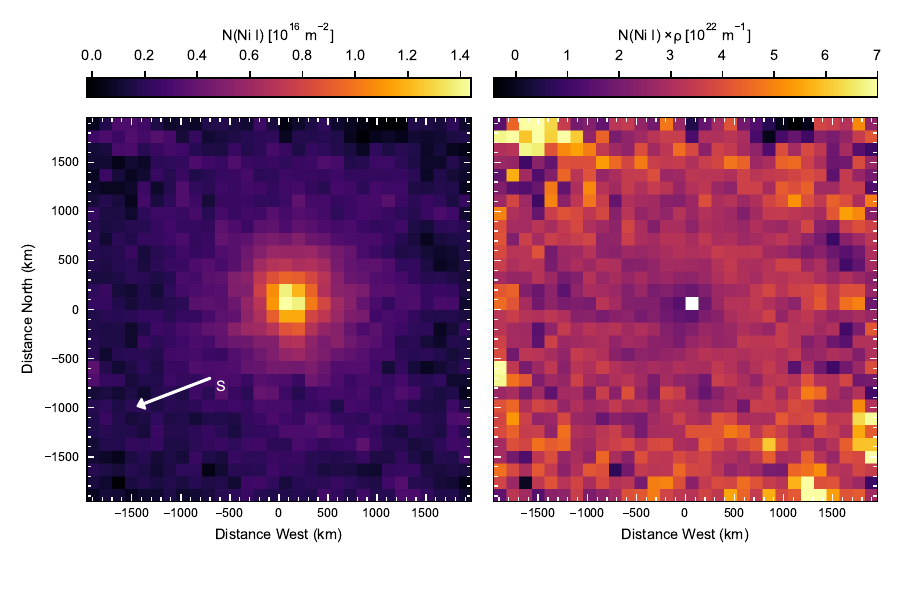}
\caption{\textbf{Left.} $N$(Ni I) map of the 3.119 $\mu$m transition in 3I/ATLAS.  \textbf{Right.} $N$(Ni I)$\times\rho$ for the 3.119 $\mu$m transition in 3I/ATLAS.
\label{fig:niMaps}}
\end{figure}

Next we considered spaxel-by-spaxel mapping of the (baseline-subtracted) 3.119 $\mu$m transition (Figure~\ref{fig:niMaps}). We calculated $\langle N \rangle=(4.95\pm0.03)\times10^{15}$ m$^{-2}$ from averaging all spaxels within a $1\farcs5$ diameter aperture, providing a range of $Q=(3.0-8.3)\times10^{24}$ \ps{} depending on the parent scale length. Our $Q(\ce{Ni})$ for both methods lie in the range reported by \cite{Belyakov2026} based on MIRI observations of 3I/ATLAS conducted on nearby dates to the NIRSpec observations reported here.

